\documentclass[aip,twocolumn,superscriptaddress,reprint]{revtex4-1}

\usepackage{amssymb}
\usepackage{amsmath}
\usepackage{graphicx}
\usepackage{natbib}
\usepackage{epstopdf}
\usepackage[version=3]{mhchem}

\begin{document}
\textheight = 63\baselineskip

\title{Large anomalous Hall effect in ferromagnetic insulator-topological insulator heterostructures}% Force line breaks with \\

\author{L. D. Alegria}
\affiliation{Department of Physics, Princeton University, Princeton, New Jersey 08544}
\author{H. Ji}
\affiliation{Department of Chemistry, Princeton University, Princeton, New Jersey 08544}
\author{N. Yao}
\affiliation{Princeton Institute for the Science and Technology of Materials, Princeton University, Princeton, New Jersey 08544}
\author{J. J. Clarke}
\affiliation{Hitachi High Technologies America, Inc., Clarksburg, Maryland 20871}
\author{R. J. Cava}
\affiliation{Department of Chemistry, Princeton University, Princeton, New Jersey 08544}
\author{J. R. Petta}
\affiliation{Department of Physics, Princeton University, Princeton, New Jersey 08544}

\date{\today}

\begin{abstract}
We demonstrate the van der Waals epitaxy of the topological insulator compound \ce{Bi2Te3} on the ferromagnetic insulator \ce{Cr2Ge2Te6}. The layers are oriented with (001)\ce{Bi2Te3}$\parallel$(001)\ce{Cr2Ge2Te6} and (110)\ce{Bi2Te3}$\parallel$(100)\ce{Cr2Ge2Te6}. Cross-sectional transmission electron microscopy indicates the formation of a sharp interface.  At low temperatures, bilayers consisting of \ce{Bi2Te3} on \ce{Cr2Ge2Te6} exhibit a large anomalous Hall effect (AHE). Tilted field studies of the AHE indicate that the easy axis lies along the c-axis of the heterostructure, consistent with magnetization measurements in bulk \ce{Cr2Ge2Te6}.  The 61~K Curie temperature of \ce{Cr2Ge2Te6} and the use of near-stoichiometric materials may lead to the development of spintronic devices based on the AHE.
\end{abstract}

%73.20.-r		Electron states at surfaces and interfaces
%75.50.Pp		Magnetic Semiconductors
%81.10-h		Methods of Crystal Growth
%85.35.Be		Quantum well devices

\maketitle

%Yang2013-cl,Hor2010-az,Checkelsky2012-uk,Haazen2012-pc

Introducing a magnetic exchange gap into the Dirac spectrum of a topological insulator (TI) surface generates circumferential, half-integer quantum Hall states.\cite{Hasan2010-oh,Yu2010-iz,Kim2012-wd,Chen2010-po}  Recently, independent experimental groups\cite{Chang2013-ht,Checkelsky2014-jx,Kou2014-ab} have detected the quantum anomalous Hall effect (QAHE) in thin films of magnetically doped topological insulators, in which the upper and lower surfaces are ferromagnetically gapped. However, the host TI materials are disordered due to the extreme concentration of magnetic dopants, which may contribute to the milli-Kelvin temperature scale at which the QAHE is observed and the reported sensitivity of the effect to fabrication details.\cite{Chang2013-ht} In contrast, introducing ferromagnetism into the TI surface by proximity to a ferromagnetic insulator (FI) leaves the TI ordered, and can break time reversal symmetry without directly alloying the TI with magnetic elements.  To date, epitaxy of TI-FI systems has been investigated using the FI materials EuS, GdN, and Y$_3$Fe$_5$O$_{12}$.\cite{Zhang2013-zh,Yang2013-cl,Kandala2013-py,Wei2013-rh}  As compared to the layered TI crystal structure, these FI compounds possess 3D structures (e.g.\ the NaCl-type structure in the case of EuS) producing radicals at a TI-FI interface.  To create a pristine interface, a FI compound is sought with a stable, lattice-matched surface.  The strong anisotropy of the \ce{Bi2Te3} crystal structure causes films to grow in the [001] direction, further constraining the problem by requiring a match to that lattice plane.

\begin{figure}[h]
\includegraphics[width=3.35 in, bb = 0 0 240 250]{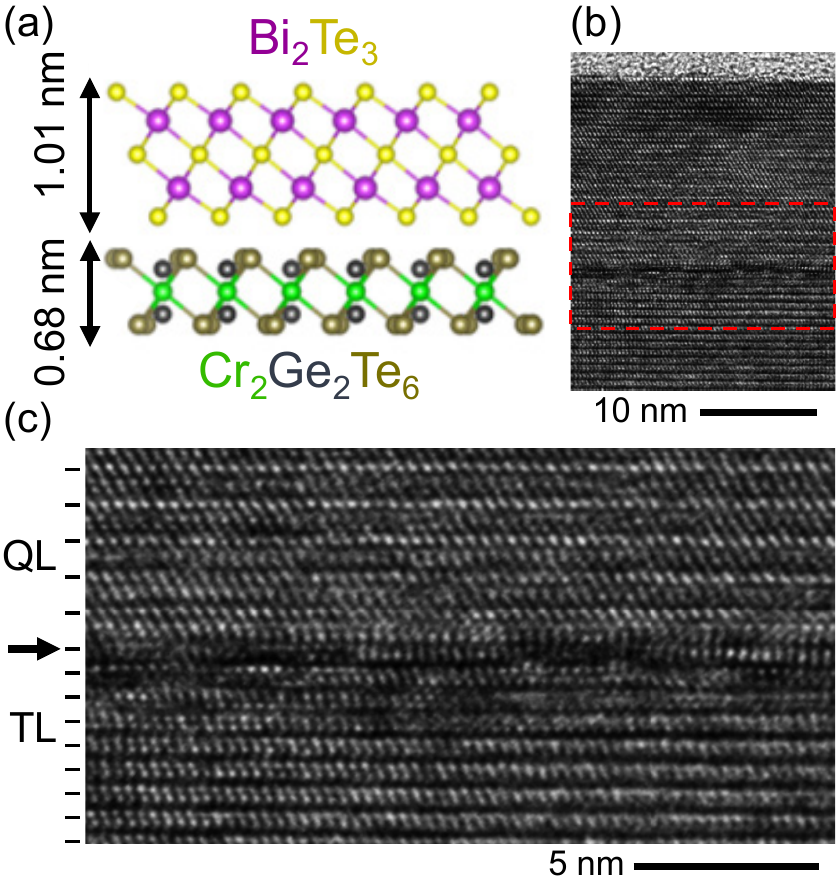}
\caption{\label{fig1} Cross-sectional TEM images. (a) The layered units of \ce{Bi2Te3} and CGT, viewed here along the basal plane, are terminated by planes of Te that make up a van der Waals interface.  (b) A TEM image of the heterostructure reveals a smooth interface. The deposited \ce{Bi2Te3} is 15 nm thick.  (c) Enlarged TEM image of the interface. The thinner (0.68 nm) CGT layers are visible in the lower portion of the image. There is an abrupt transition (marked by the arrow) to the 1.01 nm thick \ce{Bi2Te3} quintuple layers in the upper half of the image.}
\end{figure}

Here we employ the ferromagnetic insulator \ce{Cr2Ge2Te6} (CGT) as a substrate for the growth of \ce{Bi2Te3}, a standard TI material.\cite{Carteaux1995-si,Ji2013-iw,Qu2010-fi}  Like \ce{Bi2Te3}, the CGT structure consists of layered units terminated with hexagonal Te planes.   While \ce{Bi2Te3} consists of quintuple layers (QL) made of five alternating layers of Te and Bi, the layered units of CGT are narrower triple layers (TL), in which two hexagonal Te planes surround a layer of Cr ions and Ge pairs.  In both crystals, van der Waals bonds join adjacent planes of Te [see Fig.~\ref{fig1}(a)].  The hexagonal lattice of Te in CGT has a lattice constant of 3.94~\AA (=1/2 the lattice constant of the full cell), which we show below to be a good match to that of the Te surface of \ce{Bi2Te3} when rotated by 30 degrees (4.39~\AA), and similarly to the Se surface of \ce{Bi2Se3} (4.14~\AA).\cite{Y_Feutelais1993-lb}  In addition to this fortuitous crystal structure, CGT has a relatively high Curie temperature of 61~K, a resistivity greater than $10^{3}$ $\Omega$-cm below 77~K, and an easy axis of magnetization that points along the c-axis in bulk crystals.\cite{Ji2013-iw}  Taken together, these factors favor the study of the \ce{Cr2Ge2Te6}-\ce{Bi2Te3} (CGT-BT) system for the introduction of ferromagnetic order in TIs. Through Hall measurements, we show that the magnetization of CGT has a large impact on electrical transport through the \ce{Bi2Te3} layer.

To form CGT-BT heterostructures, we first grow large single crystals of CGT.\cite{Ji2013-iw} We mix high purity Cr (99.99\%), Ge (99.999\%), and Te (99.999\%) in a molar ratio of 2:6:36, where the Ge and Te excess acts as flux.  The mixture is heated to 700~$^\circ$C for 20 days and cooled to 500~$^\circ$C over 36 hours, followed by centrifugation to separate the CGT crystals from the flux. The CGT crystals are then annealed in vacuum to remove any residual Te flux.  Flat single crystals with sizes ranging from 0.5 to 3 mm can be manually isolated from the product.  X-ray diffraction, transport measurements, and magnetization measurements confirm that the samples are single crystal (with space group R$\bar{3}$), insulating ($E_g$ = 0.2 eV), and ferromagnetic ($T_C$ = 61~K).\cite{Carteaux1995-si,Ji2013-iw}

We grow the \ce{Bi2Te3} layer using metal organic chemical vapor deposition (MOCVD).\cite{Alegria2012-il} CGT is prepared for epitaxy using two different methods. We either adhere $\sim$ 500 $\mu$m CGT squares to a \ce{SiO2} carrier chip using polyimide or exfoliate mesoscopic crystals of CGT onto a \ce{SiO2} chip using the scotch tape method.\cite{Novoselov2004-qd}  In both cases, the process prepares a freshly cleaved CGT surface prior to \ce{Bi2Te3} deposition. To initiate growth of \ce{Bi2Te3}, the CGT substrate is heated to 280~$^\circ$C in a 600 sccm flow of \ce{H2} (6N) at 100 Torr.  Tellurium and bismuth are introduced into the growth chamber by flowing hydrogen gas through diisopropyl telluride (DiPTe) and trimethyl bismuth (TMBi) bubblers, resulting in partial pressures $p_{\text{TMBi}}$ = 7$\times10^{-6}$  atm and $p_{\text{DiPTe}}$ = 2.8$\times10^{-5}$ atm. The precursors thermally decompose at the sample, producing atomic Bi and Te. After 2700 s, the TMBi flow is stopped, terminating growth, and the sample is cooled to 125~$^\circ$C under continued DiPTe flow.

Cross-sectional transmission electron microscope (TEM) imaging allows for a direct characterization of the heterostructure (see Fig.~\ref{fig1}). In cross-section, the layered units of the two crystal structures are directly visible, indicating that the Te layers are stacked along the growth direction. The TEM images displayed in Figs.~\ref{fig1}(b--c) show that the interface between the two structures occurs over a distance of less than 1 nm. We attribute the high quality interface to the similarity of the two crystal structures.

\begin{figure}
\includegraphics[width=3.35 in, bb = 0 0 240 250]{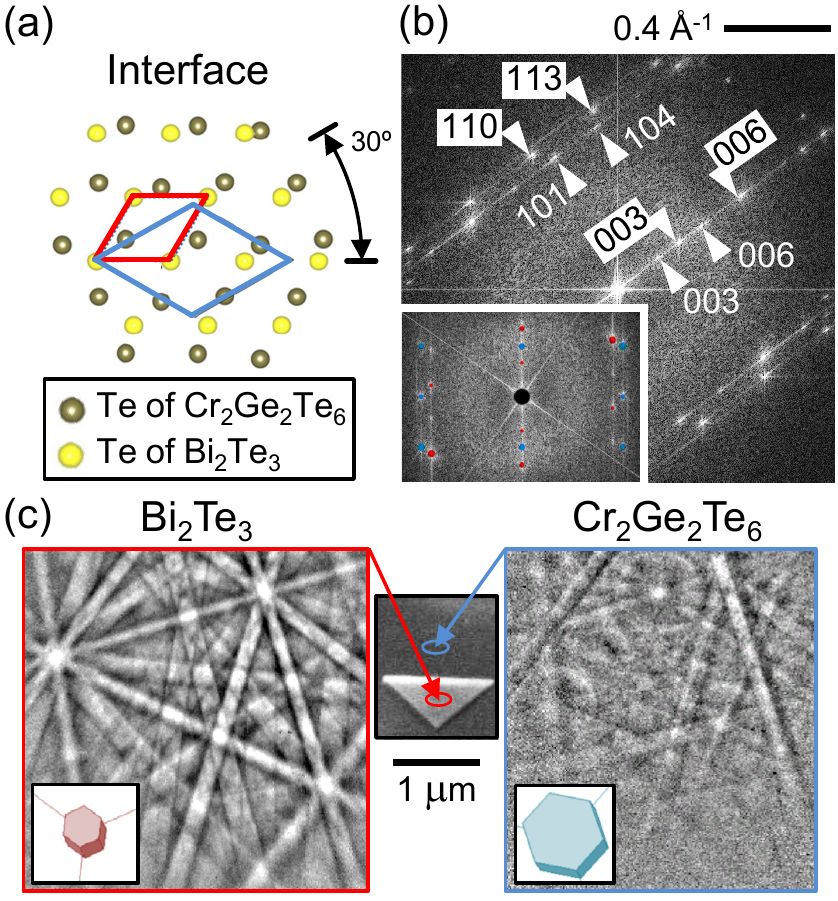}
\caption{\label{fig2} In-plane crystallographic orientation.  (a)  The interfacial hexagonal Te planes, viewed along the $c$-axis, align when the conventional unit cells of \ce{Bi2Te3} and CGT are rotated by 30$^\circ$ in-plane, i.e.\ when \ce{Bi2Te3}(110)$\parallel$CGT(100).  The boundaries of the \ce{Bi2Te3} and CGT unit cells are red and blue, respectively.  (b) Electron diffraction pattern of the cross-section sample of Fig.~\ref{fig1}, with \ce{Bi2Te3} reflections indexed with white numbers and CGT with black numbers.  In the inset, the simulated diffraction pattern for the (110)$\parallel$(100) orientation of \ce{Bi2Te3} (red) and CGT (blue) is overlaid on the data.  (c) The orientation is confirmed via electron backscatter diffraction from $\sim$200 nm spots collected from a \ce{Bi2Te3} island or the surrounding \ce{Cr2Ge2Te6} substrate. The results from automatic indexing are shown as insets. The unit cells differ by a 30$^\circ$ in-plane rotation.}
\end{figure}

Electron diffraction is used to determine the relative orientation of the \ce{Bi2Te3} and CGT layers. The indexed diffraction pattern is shown in Fig.~\ref{fig2}(b). By comparing with simulated diffraction patterns, we confirm precise alignment of the \ce{Bi2Te3} and CGT (001) axes. Additionally, \ce{Bi2Te3} (110) is parallel to CGT (100).  As illustrated in Fig.~\ref{fig2}(a), this orientation corresponds to an alignment of the interfacial Te planes.  Although TEM demonstrates the alignment locally, electron backscatter diffraction (EBSD) can be used to compare the \ce{Bi2Te3} orientation across the entire film surface.  An illustration of the method is shown in Fig.~\ref{fig2}(c), where EBSD patterns from a micron-sized \ce{Bi2Te3} island and the surrounding CGT are automatically indexed, showing a 30$^\circ$ relative rotation of the unit cells, consistent with the TEM diffraction results.  Similar analysis at many points on the \ce{Bi2Te3} film surface shows this registry to be a global feature of the CGT-BT films.

Transport measurements are performed on the CGT-BT heterostructures. We electrically contact two types of heterostructures: (a) macroscopic \ce{Bi2Te3} films grown on squares of CGT $\sim$ 500 $\mu$m wide and $\sim$ 10 $\mu$m thick (similar to those studied with TEM) and contacted with silver paint following the van der Pauw method, and (b) mesoscopic CGT-BT bilayers formed by exfoliation of CGT on a \ce{SiO2} substrate followed by \ce{Bi2Te3} deposition, which produces a $\sim$ 150 nm thick layer of \ce{Bi2Te3} on a $\sim$ 50 nm thick CGT layer, with lateral dimensions of tens of microns.  We use electron beam lithography (EBL) to define six Ti/Au contacts to each mesoscopic sample for measurements of the longitudinal and Hall resistance. Control devices consisting of EBL-contacted exfoliated CGT on \ce{SiO2} and MOCVD \ce{Bi2Te3} on \ce{SiO2} are studied in parallel with the CGT-BT heterostructures. The \ce{Bi2Te3} control samples were fabricated from \ce{Bi2Te3} islands that grew near the CGT-BT platelets (on the same chip) and measured in the same cryostat to exclude any extrinsic origin to the AHE measured in the CGT-BT samples.

\begin{figure}
\includegraphics[width=3.35 in, bb = 0 0 240 250]{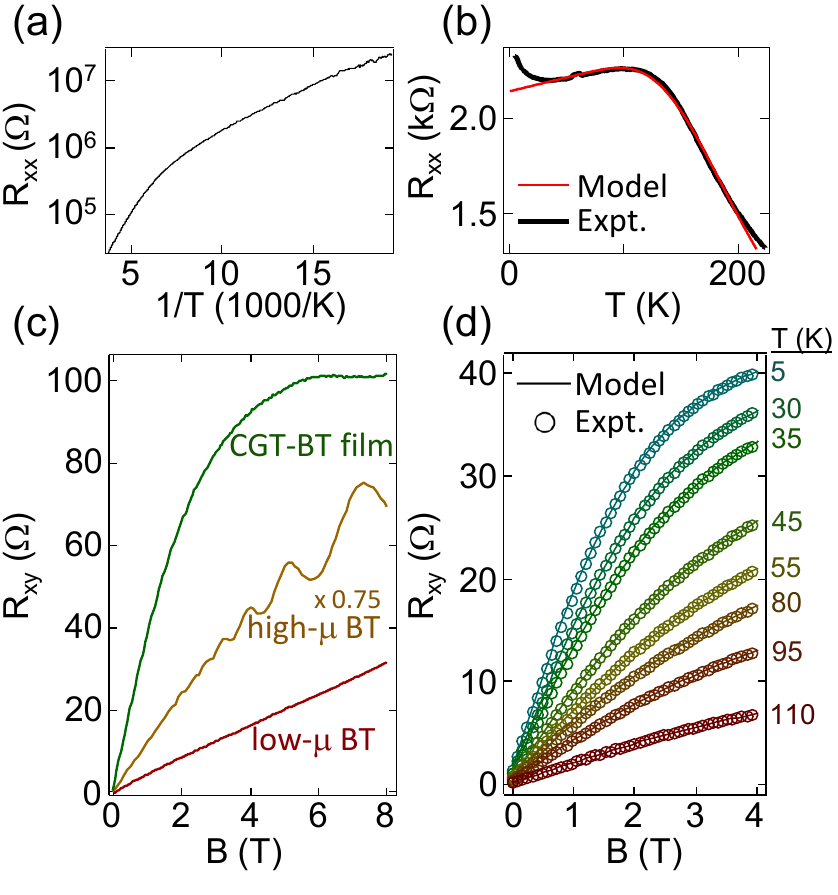}
\caption{\label{fig3} Transport in control samples and CGT-BT devices. (a) An Arhennius plot of the longitudinal resistance $R_{xx}$ of a CGT control sample exhibits a strongly insulating character with $E_g$ = 0.2 eV from the high-temperature linear region. (b) $R_{xx}(T)$ of a macroscopic CGT-BT film conforms to Eqn.~\ref{tempdep}. (c) The 4.2~K Hall resistances, $R_{xy}$, of control samples of low mobility \ce{Bi2Te3} are typically linear (low-$\mu$ BT) or, in high mobility samples, linear with superimposed Shubnikov-de Haas oscillations (high-$\mu$ BT). In comparison, the CGT-BT heterostructures show a nonlinearity that saturates at large magnetic fields.  (d) The nonlinearity observed at low temperatures gradually crosses over to a linear dependence at high temperatures. Experimental data in this panel are from sample A and the data are fit using a two-band model (Eq.~\ref{twobandR}).}
\end{figure}

Longitudinal resistance $R_{xx}$ is measured as a function of temperature to distinguish metallic (insulating) contributions from \ce{Bi2Te3} (CGT). Whereas all \ce{Bi2Te3} control samples studied show metallic $R_{xx}(T)$ characteristics, the resistances of the CGT control samples (pre- and post-EBL) show thermally activated behavior with an energy gap of 0.2 eV, as shown in Fig.~\ref{fig3}(a). For the CGT-BT bilayer, we construct a parallel resistor model by summing the \ce{Bi2Te3} and CGT conductivities:
\begin{equation}
\label{tempdep}
R_\text{CGT-BT}(T) = [(R_\text{0,BT} + a T)^{-1} + (R_\text{0,CGT} e^\frac{E_g}{2 k_B T})^{-1}]^{-1}.
\end{equation}

From the sample of Fig.~\ref{fig3}(b), we extract $R_\text{0,BT} = 2140$~$\Omega$, $a$ = 1.4~$\Omega$/K, $R_\text{0,CGT} = 65$~$\Omega$, and gap energy $E_g$ = 0.14~eV.  The corresponding room temperature resistivities of the two layers of this sample (width = 200, length = 500 um, \ce{Bi2Te3} thickness = 30 nm, CGT thickness = 10 $\mu$m) are $\rho_\text{CGT}$ = 40 m$\Omega$-cm, $\rho_\text{\ce{BT}}$ = 0.3 m$\Omega$-cm, in agreement with typical values for both materials in the literature.\cite{Carteaux1995-si,Qu2010-fi} The model gives a \ce{Bi2Te3} conductivity that is more than 1000 times greater than CGT at temperatures below 78~K in this sample. All other samples have thinner CGT and therefore smaller contributions of the CGT to the conductivity.  At 4.2~K, the three \ce{Bi2Te3} control samples show resistivities ranging from 0.40 -- 1.1 m$\Omega$-cm, while the five CGT-BT samples show resistivities in the range 0.3 - 6.2 m$\Omega$-cm, with a mean resistivity 2 m$\Omega$-cm.  The similarity of the resistivity of the CGT-BT bilayers to \ce{Bi2Te3} control samples fabricated on SiO$_2$ indicates that transport in the CGT-BT samples occurs primarily in the \ce{Bi2Te3} layer.

\begin{figure}
\includegraphics[width=3.35 in, bb = 0 0 240 180]{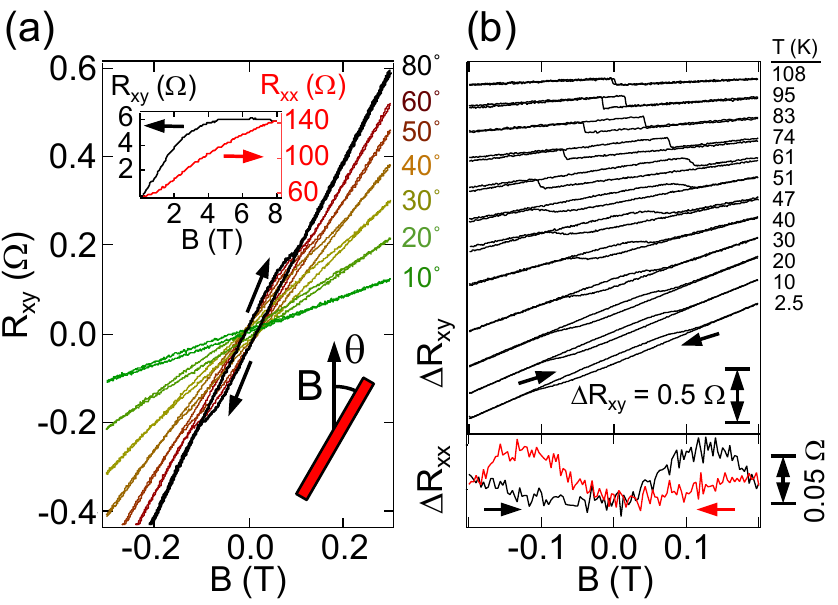}
\caption{\label{fig4} Hysteresis at low magnetic fields (sample B). (a) A hysteretic AHE is visible at low magnetic fields and superimposed on the dominant nonlinear magnetoresistance (inset). The AHE is observed in all CGT-BT samples and absent in measurements of control \ce{Bi2Te3} Hall bars.  As the sample is tilted parallel to the field, the coercive field of the hysteresis loop increases.  (b)  The hysteresis measured as a function of temperature (with field applied perpendicular to the sample). The non-monotonic dependence of the hysteresis loop size with temperature is observed in all CGT-BT samples. The hysteresis loops have been offset for clarity. (lower panel) The magnetoresistance also exhibits hysteresis as a function of perpendicularly applied field, with peaks at the coercive field. $R_{xx}$ data are from sample C at $T$ = 0.4~K.}
\end{figure}

We study magnetotransport in one CGT-BT film and in four mesoscopic CGT-BT samples, with similar results in all cases. As shown in Fig.~\ref{fig3}(c), the \ce{Bi2Te3} control samples (and \ce{Bi2Te3} samples generally) show either a linear Hall resistance ($R_{xy}$) or a linear Hall resistance with superimposed Shubnikov de Haas (SdH) oscillations in the case of high mobility samples.\cite{Checkelsky2011-hj,Qu2010-fi}  In contrast, the heterostructures all show a nonlinear Hall resistance, but without large oscillations.  The nonlinearity decreases with increasing temperature, as shown in Fig.~\ref{fig3}(d).  Whereas the \ce{Bi2Te3} control samples show sharp antilocalization features, the CGT-BT magnetoresistance $R_{xx}(B)$ does not.\cite{Alegria2012-il}  The inset of Fig.~\ref{fig4}(a) shows a typical example of $R_{xx}(B)$. Strong magnetic dephasing may suppress localization effects in the bilayer samples.\cite{Bergmann1984-ni}

In addition to the high field nonlinearity, the CGT-BT heterostructures exhibit a hysteretic AHE at low fields, as shown in Fig.~\ref{fig4}. The hysteresis is absent in all control samples consisting only of CGT or \ce{Bi2Te3}. The hysteresis observed in the CGT-BT samples is suggestive of a ferromagnetic origin.  As shown in Fig.~\ref{fig4}(b), the magnetoresistance, $R_{xx}$, also displays hysteresis, with peaks at the coercive fields. Tilted field measurements show that the coercive field of the hysteresis loop increases and the background Hall contribution falls off linearly as the field is rotated into the plane of the CGT-BT interface. The dependence is consistent with the easy axis of the ferromagnet pointing along the $c$-axis, as observed in several layered ferromagnets.\cite{Checkelsky2008-he,Ji2013-iw,Lee2014-ac}

The coercive field of the hysteresis observed here, which is $\sim$ $0.1$~T at low temperatures, is similar to the field required for saturation of bulk CGT (0.26~T).  However, bulk measurements found the width of the magnetization loop to be narrower than 0.01~T, while we observe a substantial hysteresis loop in the AHE of CGT-BT heterostructures.  The apparent hardening of the magnetization is consistent with recent measurements of \ce{EuS}-\ce{Bi2Se3} heterostructures, where the authors attributed the increase in the observed perpendicular anisotropy to exchange coupling between the TI and FI layers.\cite{Wei2013-rh}  The persistence of the hysteresis to $\sim$ 110~K, which is considerably above the known Curie temperature of bulk CGT (61~K), is also surprising.

We consider two origins of the observed high field nonlinear Hall effect. The first model considers the AHE, with the nonlinearity due to a magnetic origin. The second picture relies on a two-band model of transport, with multiple conduction channels of differing mobility.  Both models result in nearly identical predictions for $R_{xy}(B)$.

The AHE, as observed in magnetic conductors, correlates strongly with the magnetization, leading to the conventional representation $R_{xy} = R_s M_z + R_0 B/\mu_0$ where $M_z$ is the out-of-plane component of the magnetization, $R_s$ is the anomalous Hall coefficient, $B$ is the external applied field, and $R_0$ is the ordinary Hall coefficient.\cite{Nagaosa2010-vm,Hor2010-az,Haazen2012-pc}  In magnetic conductors, nonlinearity originates in $M(B)$ which often exceeds the linear component ($R_s \gg R_0$) to the extent that $R_{xy}$ appears to saturate at accessible fields, as in Fig.\ \ref{fig3}(c).\cite{Nagaosa2010-vm}  Studies of \ce{EuS}-\ce{Bi2Se3} ascribe a similar, but smaller ($R_s \ll R_0$) Hall nonlinearity to a ferromagnetic AHE, since it correlates clearly with the Brillouin dependence of the magnetization of \ce{EuS}.\cite{Wei2013-rh}  In contrast, the nonlinear Hall effect in CGT-BT does not reflect the typical $M(B)$ of CGT. Magnetization measurements show that CGT saturates at fields below 0.3~T, while here we observe saturation at $\sim$~6~T.  However, it is possible that ferromagnetic order within the CGT is strongly modified near the interface, where it would contribute most to the Hall effect.

Alternatively, the presence of parallel conduction channels with differing mobilities but comparable overall conductance will produce a similar nonlinearity, as observed in many heterojunction systems.\cite{Kim2010-mx,Qu2010-fi,Yang2014-uq}  The CGT bulk is insulating at low temperatures, so any conduction channels lie in the \ce{Bi2Te3} bulk or at the CGT-BT interface.  In the two-band model, the measured Hall resistance depends on the magnetic field as

\begin{equation}
\label{twobandR}
R_{xy}(B) = \left(\frac{B}{e}\right) \frac{(n_1 \mu_1^2 + n_2 \mu_2^2)+ (B \mu_1\mu_2)^2(n_1+n_2)}{(n_1 \mu_1 + n_2 \mu_2)^2+(B \mu_1\mu_2)^2(n_1+n_2)^2},
\end{equation}

\noindent where $n_i$ are the two-dimensional carrier densities, $\mu_i$ are the mobilities, and $e$ is the carrier charge.\cite{Yang2014-uq}  This model nicely fits the data [see fits plotted in Fig.~\ref{fig3}(c)], and implies a high mobility, low density (HMLD) channel and a low mobility, high density (LMHD) channel of the same sign (n-type) in all samples.  Due to variations in film thickness and roughness, $n_i$ and $\mu_i$ vary within one order of magnitude from sample to sample, and we consider the mean values for discussion.  At 4.2~K, the mean 2D carrier densities are $n_{\text{LMHD}} = 5 \times 10^{15}$ cm$^{-2}$, $n_{\text{HMLD}} = 6 \times 10^{13}$ cm$^{-2}$ and the mean mobilities are $\mu_{\text{LMHD}} = 200 $ cm$^2$/V~s, $\mu_{\text{HMLD}} = 7,000 $ cm$^2$/V~s.  The samples with the highest $\mu_{\text{HMLD}}$ obtained from the fits (40,000 cm$^2$/V~s) far exceed the highest known mobility for \ce{Bi2Te3}. However, large SdH oscillations are not observed in any of the CGT-BT samples.

One of the channels is consistent with bulk transport through the \ce{Bi2Te3} layer.  The fit parameters of the LMHD channel are similar to those extracted from the \ce{Bi2Te3} control samples.  The mean carrier density of the LMHD channel is  $n_{\text{3D}} = 3 \times 10^{20}$ cm$^{-3}$ a value typical of strongly doped \ce{Bi2Te3} and close to the control samples $n_{\text{3D}} = 4 \times 10^{19}$ cm$^{-3}$.  The mobility of this channel (200 cm$^2$/V~s) is also similar to the control samples (400 cm$^2$/V~s). While the LMHD channel is consistent with bulk \ce{Bi2Te3} conduction, the second high-mobility channel suggested by the two-band model could only originate from the interface between the two materials, since it is absent in either type of control sample.  In principle, a HMLD channel could be present at the interface, but this does not explain the noted absence of SdH oscillations.

We demonstrate ferromagnetic coupling in a TI-FI heterostructure with a high level of crystalline order.  To date, hysteretic AHE has not been observed in TI-FI heterostructures.  The large magnitude of both the hysteretic AHE and the high field nonlinearity in this system make this an exciting development for realizing more complex experiments based on TIs.  The hysteretic AHE reflects the bulk properties CGT, while the high-field nonlinearity implies that either an additional type of magnetic order impacts transport through the \ce{Bi2Te3}, or that a high mobility state arises at the interface.  Tuning the Fermi energy of the \ce{Bi2Te3} by applying a gate potential through the CGT layer would do much to clarify the properties of the AHE and reduce bulk conduction in the \ce{Bi2Te3}.  It is likely that heterostructures with more insulating \ce{Bi2Te3} and therefore greater sensitivity to the interface, will display still further enhanced AHE.

We thank Tian Liang and Phuan Ong for valuable discussions.  Supported by the Packard Foundation, the National Science Foundation through the Princeton Center for Complex Materials (DMR-0819860), and the Eric and Wendy Schmidt Transformative Technology Fund.

\end{document}